\documentclass[aps,prl,showpacs,superscriptaddress,twocolumn]{revtex4}
\usepackage{graphicx}
\usepackage{amssymb}
\begin{document}
\title{Supercurrents through half-metallic ferromagnetic CrO$_2$ revisited}
\author{M.S. Anwar}
\affiliation{Kamerlingh Onnes Laboratorium, Leiden University, The Netherlands}
\altaffiliation[On leave from:~]{Dept. of Physics, University of Engineering and
Technology, Lahore-54890, Pakistan.}
\author{M. Hesselberth} \affiliation{Kamerlingh
Onnes Laboratorium, Leiden University, The Netherlands}
\author{M. Porcu}
\affiliation{National Centre for High Resolution Microscopy, Kavli Institute for
Nanoscience, Technical University of Delft, The Netherlands}
\author{J. Aarts}
\email[E-mail:]{aarts@physics.leidenuniv.nl} \affiliation{Kamerlingh Onnes
Laboratorium, Leiden University, The Netherlands}
\date{\today}
\begin{abstract}
We report on the observation of a supercurrent through the half metallic ferromagnet
CrO$_2$ grown on hexagonal Al$_2$O$_3$ (sapphire). The current was observed to flow
over a distance of 700~nm between two superconducting amorphous Mo$_{70}$Ge$_{30}$
electrodes which were deposited on the CrO$_2$ film. The critical current $I_c$
increases as function of decreasing temperature. Upon applying an in-plane magnetic
field, $I_c$ goes through a maximum at the rather high field of 80~mT. We believe
this to be a long range proximity effect in the ferromagnet, carried by
odd-frequency pairing correlations.
\end{abstract}
 \maketitle

\begin{figure}
\includegraphics[width=8.6cm]{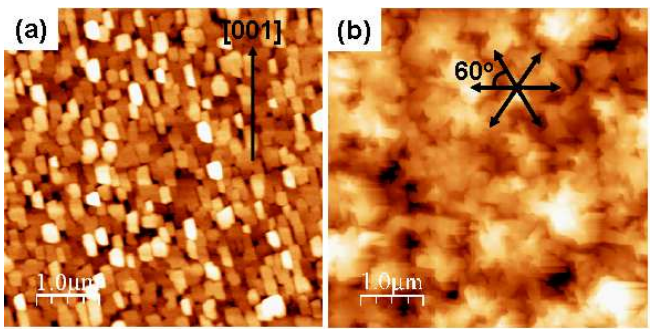}
\caption{(Color online) Atomic Force Microscopy images of CrO$_2$ films grown on (a)
a TiO$_2$ substrate; the (001) axis of the substrate is indicated; (b) an
Al$_2$O$_3$ substrate. The different directions of the crystallites which are found
in the image are indicated with the arrows and seen to make angles of 60$^{\circ}$.}
\label{CrO2sup-fig1}
\end{figure}

\begin{figure} [t]
\includegraphics[width=9cm]{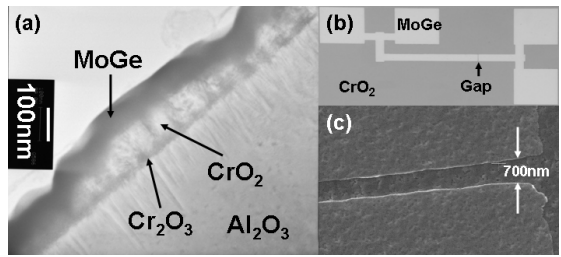}
\caption{(a) Transmission Electron Microscopy image of a CrO$_2$ film grown on
Al$_2$O$_3$. Visible are the substrate, the Cr$_2$O$_3$ seed layer, the CrO$_2$
layer, and the a-Mo$_{70}$Ge$_{30}$ layer. (b) Layout of the device structure with
four current / voltage contacts. The width of the electrodes is 30~$\mu$m. (c)
Scanning electron Microscopy image of the gap between the two electrodes, made by
lift-off. } \label{CrO2sup-fig2}
\end{figure}

\begin{figure}[t]
\includegraphics[width=6cm]{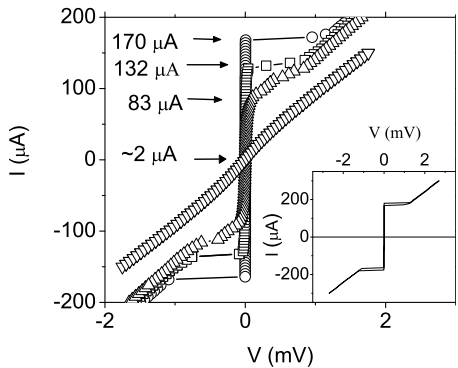}
\caption{Current ($I$) versus Voltage ($V$) measurements for device A at 2~K,
3.15~K, 4~K and 6~K. The values of the critical current are indicated. The inset
shows an I-V characteristic for device B.} \label{CrO2sup-fig3}
\end{figure}

\begin{figure}[t]
\includegraphics[width=6cm]{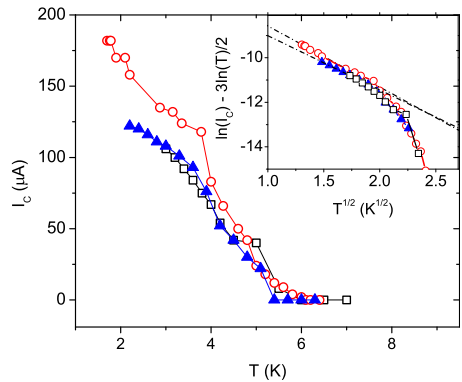}
\caption{(Color online) Critical current $I_c$ versus temperature $T$ for devices A
($\bigcirc$), B ($\triangle$), C ($\Box$). The inset shows ln$I_c$ - 3/2~ln$T$
versus $T^{1/2}$. Dashed lines correspond to a Thouless energy of 72 (91)~$\mu$V for
device A (B,C).} \label{CrO2sup-fig4}
\end{figure}

\begin{figure}[t]
\includegraphics[width=6cm]{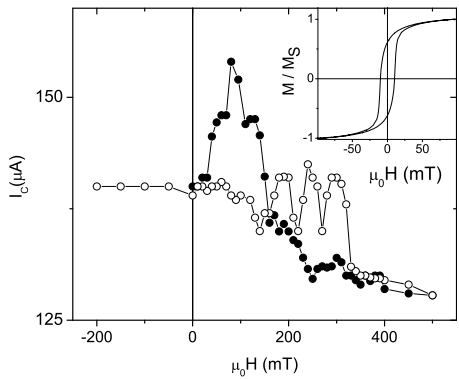}
\caption{Dependence of the critical current $I_c$ on a magnetic field in the plane
of the junction, perpendicular to the long axis of the electrodes. The insert show
the magnetization $M$ normalized on the saturation magnetization $M_s$ as function
of an in-plane applied field for a CrO$_2$ film grown on sapphire.}
\label{CrO2sup-fig5}
\end{figure}

Conventional superconductivity is generally assumed to decay rapidly inside a
ferromagnet since the spin singlet Cooper pairs are broken up by the exchange field
$h_{ex}$ when the pair diffuses through the superconductor (S)/ferromagnet (F)
interface. In the dirty limit, the decay length $\xi_F \propto 1/\sqrt{h_{ex}}$ is
no more than 10~nm even for a weak ferromagnet. Another type of pair correlations is
also possible in the magnet, based on spin triplets. These correlations have orbital
$p$-symmetry conformal to the Pauli principle \cite{eshrig03}, are only sustained in
clean systems, and can lead to a long range proximity (LRP) effect. However, it was
argued that, under the principle of odd-frequency pairing, even $s$-symmetry is
possible \cite{bergeret01,kadigrobov01}. Such correlations can even exist in dirty
systems, and would also be long ranged in the ferromagnet. To produce odd frequency
triplets in the magnet, two mechanisms were proposed. One is that the singlets on
the S-side of the interface sample an inhomogeneous magnetization on the F-side
\cite{bergeret01,kadigrobov01}. The other is a two-step process in which differences
in spin scattering at the S/F interface first cause spin mixing in the Cooper pair
singlet, leading to $m=0$ triplet correlations. Magnetic disorder then can produce
the required spin triplet component \cite{eshrig08}. The latter mechanism can be
particularly relevant when the magnet is fully spin-polarized (a so-called
half-metallic ferromagnet), since in that case triplet correlations cannot be broken
by spin flip scattering. The decay length is then set by thermal dephasing
as in the case of normal metals, and can become very long ranged. \\
Experimental indications for the LRP effect do exist. Sosnin {\it et al.}, using an
Andreev interferometer geometry, reported supercurrents flowing in ferromagnetic Ho
wires with lengths up to 150~nm \cite{sosnin06}; around the same time, Keizer {\it
et al.} reported supercurrents induced in the half-metallic ferromagnet CrO$_2$ when
superconducting electrodes of NbTiN were placed on the unstructured film, with
separations up to 1~$\mu$m \cite{keizer06}. Even for normal metals this can be
considered a very long range. Very recently, Khaire {\it et al.} prepared Josephson
junctions with ferromagnetic Cobalt, and found no decay of the value of the
Josephson current up to a thickness of 30~nm of the Co layer
\cite{khaire09}. \\
Neither Ho nor Co are fully spin polarized, and the triplet decay will be set by the
spin diffusion length, which is 100~nm (order of magnitude) in both materials. That
makes the CrO$_2$ case with its significantly larger decay length of special
interest, but here the issue of reproducibility has hampered progress. The original
report mentioned large variations in the magnitude of the critical (super)current
$I_c$ between different samples, while not all samples showed the effect
\cite{keizer06}; no other reports on experiments with CrO$_2$ have been published.
In this Letter we report new observations of supercurrents in CrO$_2$. The devices
we studied are different from the earlier ones in various aspects. In particular, we
have grown CrO$_2$ films on Al$_2$O$_3$ (sapphire) rather than on TiO$_2$, which
leads to significant differences in film morphology; and the superconducting
contacts are made from amorphous (a-)Mo$_{70}$Ge$_{30}$, rather than from NbTiN.
Again we find significant values for $I_c$ even at a separation of about 1~$\mu$m
between the electrodes, and only small sensitivity to applied magnetic fields up to 0.5~T. \\
A special issue in the device preparation lies in the growth of CrO$_2$ films. Bulk
CrO$_2$ is a metastable phase and is synthesized at high oxygen pressures.
Low-pressure film growth techniques such as sputtering, pulsed laser deposition, or
molecular beam epitaxy therefore cannot be used. Still, high-quality films can be
grown by chemical vapor deposition (CVD) at ambient pressure. For this a precursor
is used (CrO$_3$), which is heated in a furnace with flowing oxygen that transports
the sublimated precursor to a substrate at an elevated temperature, where it
decomposes and forms CrO$_2$. The method only works well, however, for substrates
with lattice parameters closely matching the $b$-axis of the tetragonal CrO$_2$ ($b$
= 0.4421~nm), such as TiO$_2$(100) (quasi-orthogonal with $b$ = 0.4447~nm) or
Al$_2$O$_3$(0001) (hexagonal with $a$ = 0.4754~nm), for which the method was
specifically demonstrated \cite{ji99,kamper87}. For our experiments, films were
grown on both types of substrates in the manner described above, with the precursor
at 260~$^{\circ}$C, substrates at 390~$^{\circ}$C, and an oxygen flow of 100~sccm.
Deposition on TiO$_2$ leads to films with a morphology widely different from films
grown on Al$_2$O$_3$, as can be seen from the images in Fig.~\ref{CrO2sup-fig1} made
by atomic force microscopy. The TiO$_2$ substrate has an almost square surface net,
and the film structure consists of (elongated) platelets as was shown earlier
\cite{miao06}. The hexagonal structure of Al$_2$O$_3$ leads to growth of
crystallites along all 6 major axes, and to considerably more surface roughness.
An important point to make is that growth on Al$_2$O$_3$ does not start as CrO$_2$,
but as Cr$_2$O$_3$, which is isostructural to the sapphire. Only after a layer of
about 40~nm has been deposited, the growing film becomes CrO$_2$ as found in
Ref.~\cite{rabe02}. Using Transmission Electron Microscopy, our finding is similar,
(Fig.~\ref{CrO2sup-fig2}a). The Cr$_2$O$_3$ layer is visible as a dark band of about
30~nm adjacent to the substrate, followed by a brighter area of about 100~nm which
is the CrO$_2$ film . The dark band on top is the a-Mo$_{70}$Ge$_{30}$ used as
superconducting contact. Also visible is the columnar structure of the film. Since
it is difficult to determine the film thickness from a calibrated deposition time,
we used the magnetic properties instead. Cr$_2$O$_3$ is an antiferromagnetic
insulator and CrO$_2$ a ferromagnetic metal with a magnetic moment of
2.0~$\mu_B/$Cr-atom ($\mu_B$ is the Bohr magneton), and the measured magnetic moment
was used to calculate the CrO$_2$ thickness. The films were characterized by
electrical transport measurements. Both specific resistance and saturation
magnetization behave as expected, with the (low temperature) residual resistivity
$\rho_0$~= 7 (10) $\mu \Omega$cm for films on TiO$_2$ (Al$_2$O$_3$). \\
The insulating nature of the substrates is an impediment in lithography. In
particular it is difficult to etch a structure into the film and then define
electrodes on the bare substrate with electron beam lithography. Instead, we made
the devices by (RF-)sputtering 60~nm of (a-)Mo$_{70}$Ge$_{30}$ superconducting
electrodes with a superconducting transition temperature T$_c \approx$ 6.5~K through
a lift-off mask onto the unstructured film. Before the deposition, the film surface
was cleaned briefly with an $O_2$ reactive ion plasma, in order to remove resist or
developer residues. Ar-ion etching was applied immediately prior to deposition, in
order to remove newly formed Cr$_2$O$_3$ on the film surface. The width of the
electrodes was about 30 $\mu$m, and the gap between the electrodes around 700~nm.
Fig.~\ref{CrO2sup-fig2}b,c show the layout of the electrodes on the film surface and
an electron microscopy image of the gap. \\
A number of devices were prepared in this way, and 3 out of roughly 10 showed a
supercurrent. We call them A,B, and C; device B was slightly different from the
other two in that it consisted of three parallel electrodes rather than one, with a
distance between electrodes of 100~$\mu$m, and the three gaps measured in parallel.
Fig.~\ref{CrO2sup-fig3} shows the current-voltage ($I$-$V$) characteristic of device
A, taken between 6~K (just below $T_c$) and 2.5~K. We observe a clear zero
resistance supercurrent branch, with a maximum value for $I_c$ of 170~$\mu$A at
2.5~K. The inset shows data for device B measured at 2~K.
From these measurements $I_c$ was determined as the first deviation from the linear
I-V characteristic around zero bias (equivalent to the peak in the derivative
$dI/dV$). The temperature dependence $I_c(T)$ is given in Fig.~\ref{CrO2sup-fig4}
for all three samples. All devices have very similar values for the critical
current, even for the case of three parallel electrodes. The behavior close to $T_c$
is concave rather than linear. In Fig.~\ref{CrO2sup-fig5} we present the effect of
applying a magnetic field $H_a$ on $I_c$ in device A at a temperature of 3~K. The
field was applied in the plane of the film, with a direction either parallel to the
long axis of the electrodes (not shown), or perpendicular to that axis. In the first
configuration we do not find effects up to 500~mT. In the second configuration we
find large changes, however. Starting from zero field, $I_c$ increases by about 10\%
and goes through a maximum around 80~mT before dropping down to a level which at
500~mT is about 10\% below the zero field value. Sweeping back, the behavior is
different, with a relatively sharp jump back to the zero field level, but no peak as
in the forward sweep. Continuing in the negative field quadrant, no structure in
$I_c(H_a)$ was found. A point to note is that the maximum lies well outside the
hysteresis loop of the magnet. The coercive fields $H_c$ are of the order of 10~mT
only (inset of Fig.~\ref{CrO2sup-fig5}). Unfortunately, the samples proved fragile
and could only be cooled down a few times before the supercurrent disappeared,
although there was no slow degradation as long as the supercurrent was present.
\\
The results are best discussed in comparison with the previous report on
supercurrents in CrO$_2$ \cite{keizer06}. Firstly, we can compare their magnitudes
by assuming that the current flows homogeneously across the bridge and through the
full thickness $d_{CrO_2}$ of the layer. In our case ($d_{CrO_2} \approx$ 100~nm,
bridge width 30~$\mu$m, current 100~$\mu$A) we find a critical current density at
2~K of about 3$\times 10^7$~[A/m$^2$]. The earlier data ($d_{CrO_2}$ = 100~nm,
bridge width 2~$\mu$m, typical current 1~mA) correspond to 5$\times 10^9$~[A/m$^2$],
and from this point of view there appears to be a large difference between the two
results. Another way of looking at them is to compare the field dependence. In
Ref.~\cite{keizer06}, a Fraunhofer pattern was detected with a distance between
maxima of about 90~mT. Assuming this equivalent to one flux quantum $\Phi_0$ in the
junction area of 310~nm $\times \; d_{CrO_2}$, a value of roughly 80~nm is found for
$d_{CrO_2}$, quite close to the nominal thickness and suggesting that the full film
thickness is partaking in the supercurrent (c.q. in the shielding from the magnetic
field). In the dataset presented here (Fig.~\ref{CrO2sup-fig5}) a Fraunhofer pattern
is not clearly visible, but there is a maximum at 80~mT followed by discontinuities
around 150~mT and 250~mT, and a small maximum at 300~mT.
Taken together, this suggests a period of 100~mT. For a junction area of 700~nm
$\times \; d_{CrO_2}$, this corresponds to $d_{CrO_2} \approx$~30~nm, which
indicates that in our case the current is not flowing through the full thickness of
the layer. The picture then emerging is that, although the results are qualitatively
the same, the growth on Al$_2$O$_3$ leads to a somewhat weaker junction. Since the
TEM picture in Fig.~\ref{CrO2sup-fig2} shows that in our devices grain boundaries
will always be in the path of the current, this actually seems a reasonable
conclusion. Another point to discuss is that the maximum in the Fraunhofer pattern
is not found at zero field. In Ref.~\cite{keizer06} this was ascribed to the finite
sample magnetization. The difficulty here is that, starting from the demagnetized
(domain) state, the maximum can be expected when magnetization saturation is
reached, which is at a significantly smaller field value. An alternative explanation
is that in the case of a triplet supercurrent the junction is expected to show a
phase shift of $\pi$ \cite{braude07,asano07}; such a $\pi$-junction would show a
minimum at zero-field rather than a maximum.
\\
There is another way to gauge the strength of the junction. According to diffusive
theory, $I_c$ for a long S-N-S junction (N a normal metal) is proportional to
$T^{3/2} exp \; (\sqrt{(2\pi k_B T) / E_{Th}})$, with $E_{Th}$ the Thouless energy
given by $(\hbar D) / L^2$; here, $D$ is the diffusion constant of the N metal and
$L$ the length of the junction \cite{wilhelm97,dubos01}. Plotting ln($I_c$) -
3/2~ln(T) versus $\sqrt{T}$ in the inset of Fig.~\ref{CrO2sup-fig4} shows that the
relation holds well at low temperatures, with values for $E_{Th}$ of 72 (91)~$\mu$V
for device A (B,C). This in turn can be used to estimate the maximum critical
current from the relation $e I_c R_N$ = 10.8~$E_{Th}$ \cite{dubos01}. The normal
resistance $R_N$ of the junction is 11~$\Omega$, which would yield a value for $I_c$
of 75~$\mu$A. Although this compares well to the measurements, the measured $R_N$
does not seem the appropriate value to use, since it is dominated by the contact
resistance. Using a typical specific resistance, measured in various films, of
10~$\mu \Omega$cm, we rather estimate the normal resistance of the junction to be
4~m$\Omega$, which would yield a value for $I_c$ of 200~mA. Although this may be an
overestimation, it indicates that the measured values for $I_c$ are actually
lower than what can be achieved. \\
Overall, the numbers suggest in several ways that the junctions are weaker than
could be, leading to a discussion of the reproducibility, including the fact that
not every device shows a supercurrent. In working devices, the current densities are
large enough to conclude that the effect is intrinsic, rather than carried by
filamentary normal metal shorts in the ferromagnetic matrix, for which also
otherwise no signs exist. Our premise is that the supercurrent is of triplet nature,
and a difficulty lies in the preparation of the 'spin-active' interface, which
should both provide the difference in spin scattering and unaligned magnetic moments
\cite{eshrig03}. Experimentally, the CrO$_2$ film surface is sensitive to oxidation
and has to be cleaned before the superconducting electrodes are deposited. The
Ar-etching will not only remove unwanted oxides, but may also damage the surface in
such a way that the required scattering or magnetization disorder is not present.
Especially the fact that device B, which consists of three parallel electrodes
rather than one, does not show a larger $I_c$, strongly suggests that the triplet
generation takes place at isolated spots under the electrodes rather than
homogeneously over their width. Another hindrance is the finite lifetime of the
devices, which is probably due to the grainy nature of the films and thermal
expansion differences between film and substrate. These may not be the only
bottlenecks, however. One common factor between the earlier experiments using
TiO$_2$ and the present ones with sapphire is that the films have more than one easy
axis of magnetization. In the case of sapphire, this is due to the strong
polycrystalline nature of the growth, particularly evident in
Fig.~\ref{CrO2sup-fig1}b. In the case of TiO$_2$, it was due to the peculiar
circumstance that strain relaxation in the film can lead to a change in the
easy-axis direction, with biaxial behavior occurring around a film thickness of
100~nm \cite{goennen07}. Apart from the experiments we report here, we have grown a
number of films on TiO$_2$, and we find that growth conditions (including substrate
cleaning prior to the growth) crucially determine whether biaxial behavior occurs.
The devices we prepared using TiO$_2$ substrates did not show supercurrents, but
those films showed uniaxial anisotropy at low temperatures, so that a true
comparison with the earlier work was not made.\\
In conclusion, we have provided new evidence that a supercurrent can flow through
the halfmetallic ferromagnet CrO$_2$ over ranges of the order of a $\mu$m. The
odd-frequency pairing scenario appears a plausible one, both from the critical
current values and from the magnetic field dependence. However, the analysis shows
that the junctions are far from perfect, and that no control as yet exists over the
preparation of the spin-active interface. This explains the difficulties in
producing reproducible experiments. \\

We thank T.M. Klapwijk, F. Czeschka, S. G\"{o}nnenwein, H. W. Zandbergen and V.
Ryazanov for useful discussions. M.S.A. acknowledges the financial support of the
Higher Education Commission (HEC) Pakistan.

\end{document}